\documentstyle[12pt]{article}
\begin{document}

\title{Flavor without Flavor Symmetry}

\author{{\bf S.M. Barr} \\ Bartol Research Institute\\
University of Delaware\\ Newark, DE 19716}

\date{BA-01-25}
\maketitle

\begin{abstract}

Non-trivial patterns of quark and lepton masses and mixings can 
arise without there being any underlying flavor symmetry that
distinguishes among the three families. Two realistic examples
are given.

\end{abstract}

\newpage

There are two outstanding features of the quark and lepton spectrum. 
First, the masses of the fermions of each kind exhibit an interfamily
hierarchy: $m_3 \gg m_2 \gg m_1$. Second, these hierarchies appear to be
nearly aligned, at least in the case of the quarks, in the sense
that $u_3$ is aligned with $d_3$, $u_2$ with $d_2$, and $u_1$ with $d_1$.
That is to say, the mixing angles are very small. There is also some
evidence that the lepton hierarchies are aligned, namely the fact that
the lepton mixing angle $U_{e3}$ is very small. But the large mixing angle
$U_{\mu 3}$ seen in atmospheric neutrino oscillations
shows that the alignment is not as good for the leptons as for the
quarks.

Attempts to explain the quark and lepton masses
are usually based on either grand unification
or flavor symmetry, or a combination of the two approaches. The way
that flavor symmetry would explain the parallel hierarchies of the
fermion masses is quite straightforward. If flavor symmetry 
distinguished the different families from each other, then
the masses of the fermions of different families would typically 
arise at different order in flavor
symmetry breaking. Moreover, the mixing of families would also be a
flavor-symmetry-breaking effect. Consequently, 
if family-symmetry-breaking effects were small, 
one might expect both a hierarchical pattern 
of masses and small mixing angles, as observed.

Turning to grand unification, one finds that grand unified gauge 
symmetries are able to
explain small mixing angles in a completely different way, which
is most readily understood by considering the minimal $SO(10)$ model.
In minimal $SO(10)$
all four Dirac mass matrices, those of the neutrinos, up quarks, down
quarks, and charged leptons (which matrices we denote henceforth as
$N$, $U$, $D$, $L$) are exactly proportional: $N = U \propto D = L$.
Obviously this means that the hierarchies in minimal $SO(10)$ are
{\it exactly} aligned, and the CKM angles vanish. 
In realistic unified models
the relation between the mass matrices is more complicated, but for
models based on $SO(10)$ and some related groups the CKM angles do
tend to be small.

The question arises whether grand unification can also explain
the other main feature of the quark and lepton spectrum 
--- the interfamily mass hierarchies ---
without flavor symmetry. If so, then it becomes
attractive to dispense with flavor symmetry altogether. By flavor
symmetry, we mean here any symmetry that distinguishes among the
three families.

If there is no flavor symmetry, then how could one explain
that some families have much larger
mass than others? A simple possibility was suggested quite a long 
time ago \cite{b80} \cite{b90}, namely
that the different families may mix differently with superheavy fermions
in real representations of the unified group.
The idea can be illustrated very simply. Suppose that we consider an
$SO(10)$ model \cite{b90} in which, in addition to the three families 
${\bf 16}_i$, $i=1,2,3$, there is a ``vectorlike" family-antifamily
pair ${\bf 16} + \overline{{\bf 16}}$. We can imagine 
a $Z_2$ parity under which the ordinary
families are odd and the vectorlike ones are even. We do not
consider this a flavor symmetry because it does not distinguish
among the three families. Suppose further that there are Higgs in the
vector and adjoint representations of $SO(10)$, ${\bf 10}_H$ and
${\bf 45}_H$. If these are also odd under the $Z_2$, then only the 
following types of renormalizable Yukawa coupling are allowed:

\begin{equation}
W_{Yukawa} = M(\overline{{\bf 16}} {\bf 16}) + 
a_i (\overline{{\bf 16}} {\bf 16}_i) {\bf 45}_H +
b_i ({\bf 16} {\bf 16}_i) {\bf 10}_H.
\end{equation}

\noindent
The mass $M$ is of the GUT scale.
The interesting point to note is that even though no symmetry has been
imposed that distinguishes among the families, two of the light families
get mass and one does not. The reason for this is that the two
Yukawa coupling vectors, $a_i$ and $b_i$, span only a two dimensional
subspace of the full three dimensional family space. To be more
concrete, one can without any loss of generality choose the axes
in family space so that $a_i = a (0,0,1)$ and $b_i = b (0, \sin \theta,
\cos \theta)$. It is then clear that the first family has no Yukawa
couplings at all and remains exactly massless. The other families get
mass through mixing, as follows. The first two terms of Eq. (1) lead to
a superheavy fermion mass term that can be written as
$\overline{{\bf 16}}(M {\bf 16} + a \langle {\bf 45}_H \rangle
{\bf 16}_3)$. One sees
that the superheavy spinor is a linear combination 
of the ${\bf 16}$ and ${\bf 16}_3$. The linear combination orthogonal to 
this is light and in fact is just the third family. In other words, 
the ${\bf 16}$ without an index actually contains in part the light 
fermions of the third family. Consequently, the third term in Eq. (1)
generates weak-scale mass terms of the form
$b \langle {\bf 10}_H \rangle {\bf 16}_3 ( \cos \theta {\bf 16}_3
+ \sin \theta {\bf 16}_2)$. That is, it generates
23, 32, and 33 elements of the light fermion
mass matrices.

To summarize what is going on, the light fermions are only able to
obtain mass through mixing with superheavy fermions. But not
all the three families are able to mix in this way, since
there are not enough superheavy fermions for them all to mix with.
Thus, perforce, an interfamily mass hierarchy results
even though all three families have exactly the same quantum numbers.

Let us examine the structure we have just described in more detail.
One may write the VEV of the adjoint Higgs field as
$\langle {\bf 45}_H \rangle = \Omega T$, where $T$ is a generator
of $SO(10)$. Then, integrating out the vectorlike fields 
$\overline{{\bf 16}} + {\bf 16}$, as shown in Fig. 1, 
one obtains the effective
operator 

\begin{equation}
\begin{array}{ccl}
W_{eff} & \cong & [a_i \langle {\bf 45}_H \rangle {\bf 16}_i]
[b_j \langle {\bf 10}_H \rangle {\bf 16}_j]/M \\ \\
& = & (a_3 T \cdot {\bf 16}_3)(b_3 {\bf 16}_3 + b_2 {\bf 16}_2) 
\langle {\bf 10}_H
\rangle (\Omega/M) \\ \\
& \propto & T_{(16_3)} {\bf 16}_3(\cos \theta {\bf 16}_3 + \sin \theta
{\bf 16}_2). \end{array}
\end{equation}

Consider now fermions of type $f$, where $f$ can be (left-handed)
up-type quarks, down-type quarks, charged leptons, or neutrinos. 
The left-handed anti-fermions are denoted $f^c$. There arises
straightforwardly from the previous equation the following effective
three-by-three mass matrix for the light fermions of type $f$:

\begin{equation}
f^c_i M_{ij} f_j \cong M_f
(f^c_1, f^c_2, f^c_3) \left( \begin{array}{ccc}
0 & 0 & 0 \\ 0 & 0 & \sin \theta T_f \\
0 & \sin \theta T_{f^c} & \cos \theta (T_f + T_{f^c})
\end{array} \right) \left( \begin{array}{c} f_1 \\ f_2 \\ f_3 
\end{array} \right).
\end{equation}

\noindent
The factor $M_f$ has one value, which we shall call $M_U$, 
for up quarks and neutrinos, and another value, which we shall call
$M_D$, for down quarks and charged leptons. The symbols $T_f$ and
$T_{f^c}$ stand for the charges of the fields $f$ and $f^c$
under the $SO(10)$ generator $T$.

There are several interesting features of this structure that have
been pointed out in earlier papers \cite{b90} \cite{bb}. 
Not only is the first family
singled out as massless despite there being no flavor symmetry that
distinguishes one family from another, but the structure naturally
accommodates a mass hierarchy between the second and third families
as well. For example, if we assume that $T_f \sim T_{f^c}$ and that
$\sin \theta$ is somewhat 
small, then $m_{f_2}/m_{f_3} \sim \frac{1}{4} \tan^2
\theta \ll 1$.

Another interesting feature of the this structure
is that it naturally explains why
the minimal $SU(5)$ relation $m_b \cong m_{\tau}$ works well
while the corresponding relation for the second family,
$m_s \cong m_{\mu}$, does not. The reason has to do with the way
the $SO(10)$ generators appear in Eq. (3). Since the same Higgs
doublet $H_d$ in ${\bf 10}_H$ couples to both $d^c d$ and $\ell^+
\ell^-$, it follows that $T_{d^c} + T_d = T_{\ell^+} + T_{\ell^-}
= - T_{H_d}$. Consequently, the 33 elements of the mass matrices for
the down quarks and charged leptons are approximately equal. 
However, the 23 and 32 elements are different for the two matrices since 
$T_d$ and $T_{d^c}$ are not equal to $T_{\ell^-}$ and $T_{\ell^+}$.

Finally, the structure incorporates automatically a sort of
Fritzschian form \cite{fritzsch} for the heavier families, with a ``texture
zero" in the 22 element. This relates the smallness of $V_{cb}$
to the smallness of $m_c/m_t$ and $m_s/m_b$. 

It is remarkable that the one effective Yukawa term given in Eq. (2)
goes so far toward providing a satisfactory framework for describing
the masses and mixings of the fermions of the second and third
families. In several earlier papers attempts were
made to construct models of the quark and lepton masses and mixings
on the basis of exactly this term. The first attempt was 
\cite{b90}, where
the second and third families were described using {\it only} the operator of 
Eq. (2). However, the resulting fits were not satisfactory. There are four
dimensionless quantities ($m_{\mu}/m_{\tau}$, $m_s/m_b$, $m_c/m_t$,
and $V_{cb}$) that had to be fit using two parameters, namely
$\theta$ and a parameter specifying the $SO(10)$ generator $T$. 
(Since $T$ must commute with the Standard Model group, it must
be a linear combination of weak hypercharge and the generator $X$
in $SU(5) \times U(1)_X$, or equivalently a linear combination 
of $B-L$ and $I_{3R}$.
Thus, only a single parameter is needed to
specify T.) It turned out that at least one quantity 
got very badly fit for all choices of parameters. In the same paper, 
a better fit was sought by extending
the model in the obvious way to the group $E_6$. One more parameter
was thereby introduced, since in $E_6$
two parameters are needed to specify the generator $T$.
It seems almost prophetic in light of recent results that the best fit
obtained in the $E_6$ model had very small $m_s$ and a large mixing
between $\mu^-_L$ and $\tau^-_L$, i.e. a large contribution
to the leptonic mixing $U_{\mu 3}$.
Unfortunately, the values of $m_s$ obtained were
{\it too} small even compared to the recent lattice results, and the 
value of the $\mu \tau$ mixing angle was not large {\it enough} to 
account for the atmospheric neutrino oscillations.

In \cite{bb} a realistic model was obtained by introducing structures
that went beyond Eq. (2) to account for the heavy two families.
It was assumed that $T = (I_{3R}) + \epsilon (B-L)$, $\epsilon \ll 1$,
which gave an
appealingly simple explanation of the Georgi-Jarlskog relation;
however, the ratio $m_c/m_t$ remained a problem, since with this
choice of $T$, the matrices in Eq. (3) give the minimal $SO(10)$
result $m_c/m_t \cong m_s/m_b$. A way of suppressing $m_c/m_t$ was
found in \cite{bb} that, although elegant, was somewhat involved.

In \cite{abb} and \cite{ab2} 
a very successful model of quark and lepton models
was constructed that also included the operator in Eq. (2).
(For a similar model see \cite{bpw}.)
However, to give a satisfactory account of the heavy two families, two
other operators in addition to that in Eq. (2) were needed. 
Nevertheless, this 
did not lead to a loss of predictivity for two reasons. First, 
the generator $T$ was fixed to be exactly $B-L$ in order to
solve the doublet-triplet splitting problem via the Dimopoulos-Wilczek
mechanism \cite{dw}, thus reducing the number of parameters by one. Second,
with $T = B-L$, the 33 elements in Eq. (3)
vanish, since $(B-L)_f + (B-L)_{f^c} = 0$, making 
the parameter $\theta$ irrelevant. 
(This necessitated, of course, that a different operator be introduced
to generate the 33 elements.)
The model of \cite{abb} and \cite{ab2} was extremely predictive and simple in
structure. One of its great successes 
was that it naturally accounted for the largeness of the atmospheric
neutrino mixing angle. However, though very simple, it made an
enormous sacrifice from the
point of view of the idea we are exploring in the present paper: it was
based on a flavor symmetry, i.e. a symmetry 
that distinguished the three families
from each other. 

To summarize the past efforts, one can say that no completely
satisfactory model exists that succeeds in explaining the flavor
structure of the quarks and leptons without a flavor symmetry.
In this paper we shall pursue this goal again. We present
two models. Both incorporate the operator in Eq. (2). The first
model is similar in spirit to the models of \cite{b90} and \cite{bb}. 
It is rather simple and has a single prediction, namely the mass
of the strange quark, which comes out smaller than the Georgi-Jarlskog
prediction and more in line with the recent lattice estimates.
The second model is very close to the model of \cite{abb} and \cite{ab2},
but is
obtained without recourse to flavor symmetry.

\vspace{0.5cm}

\noindent
{\bf Model 1.} A realistic variant of the models of \cite{abb} and \cite{ab2}
can be constructed in a simple fashion. Consider an $SO(10)$ model
with the following Yukawa superpotential;

\begin{equation}
\begin{array}{ccl}
W_{Yukawa} & = & M(\overline{{\bf 16}} {\bf 16}) 
+ a_i (\overline{{\bf 16}} {\bf 16}_i) {\bf 1}_H \\ \\
& + &  M' (\overline{{\bf 16}}' {\bf 16}') + 
b_i ( \overline{{\bf 16}}' {\bf 16}_i) {\bf 45}_H \\ \\
& + & c ( {\bf 16} {\bf 16}) {\bf 10}_H^{up} +
d ( {\bf 16}' {\bf 16}) {\bf 10}_H^{down}.
\end{array}
\end{equation}

\noindent
The vector Higgs fields ${\bf 10}_H^{up}$ and ${\bf 10}_H^{down}$
are supposed, respectively, to obtain VEVs in their $Y/2 = +1/2$
and $Y/2 = -1/2$ components. Thus, the former gives mass only to
up quarks and neutrinos, while the latter gives mass to down quarks
and charged leptons. 

The structure that emerges from these terms can be understood
readily. As before, we can write $\langle {\bf 45}_H \rangle 
= \Omega T$, where $T$ is an $SO(10)$ generator, which we choose
to parametrize as $T =  2 I_{3R} + 3d (B-L)$. (The parameter called $d$
here is the same up to a normalization as the parameter called 
$\epsilon$ in \cite{bb}.) Without
loss of generality a basis in family space can be chosen so that
the Yukawa coupling vectors take the form $a_i = a (0, 0, 1)$
and $b_i = b ( 0, \sin \theta, \cos \theta)$.
The combination of the two terms involving the 
$\overline{{\bf 16}}$, namely the term with $M$ and the term with
${\bf 1}_H$, cause the fields ${\bf 16}$ and ${\bf 16}_3$ to mix with
each other. That means that the field ${\bf 16}$ is not purely
superheavy, but also contains an admixture of the fields of the third family.
Similarly, the two terms involving the $\overline{{\bf 16}}'$, namely the
term with $M'$ and the term with ${\bf 45}_H$, cause the ${\bf 16}'$
and the linear combination $b_i {\bf 16}_i = 
b(\cos \theta {\bf 16}_3 + \sin \theta
{\bf 16}_2)$ to mix with each other. Thus, the ${\bf 16}'$ is also not
purely superheavy, but contains an admixture of
the fields of the second and third
families in a proportion that depends on the angle $\theta$.

Given these facts, one sees that the term $({\bf 16} {\bf 16})
{\bf 10}_H^{up}$ contributes only to the 33 element of the 
up quark mass matrix $U$. Similarly, the term $({\bf 16}' {\bf 16})
{\bf 10}_H^{down}$ contributes to the 23, 32, and 33 elements of
the mass matrix of the down quarks, $D$, and the mass matrix of the charged
leptons, $L$. At this stage, then, one has that the charm quark is
massless. It, like the fermions of the first family, is
supposed to get mass from other smaller terms. This is not
unreasonable, in light of the fact that $m_c/m_t$ is an
order of magnitude smaller than $m_s/m_b$ and $m_{\mu}/m_{\tau}$.

From the foregoing one can write down the following expressions
for the mass matrices $D$ and $L$:

\begin{equation}
D = \left( \begin{array}{ccc}
0 & 0 & 0 \\
0 & 0 & (1-d) \sin \theta \\ 0 & d \sin \theta & \cos \theta 
\end{array} \right) M_D, \;\;\;\; L = \left( \begin{array}{ccc}
0 & 0 & 0 \\ 0 & 0 & (1 + 3d) \sin \theta \\ 0 & -3d \sin \theta
& \cos \theta \end{array} \right) M_D.
\end{equation}

\noindent
In this model, two parameters, $d$ and $\theta$ are available
to predict three dimensionless quantities, $V_{cb}$, $m_{\mu}/m_{\tau}$,
and $m_s/m_b$. There is therefore one prediction, which can be taken to
be for the strange quark mass. For brevity, let us
define $t \equiv \tan \theta$, $v \equiv V_{cb}$, and $3 \ell \equiv
m_{\mu}/m_{\tau}$. Then, to the leading two orders in small quantities
one can write

\begin{equation}
v \equiv V_{cb} =  \frac{dt}{1 + [(1-d)^2 - d^2]t^2}.
\end{equation}

\begin{equation}
\ell \equiv \frac{1}{3} \frac{m_{\mu}}{m_{\tau}} = 
\frac{d(1+3d) t^2}{1 + [(1 + 3d)^2 + (3d)^2]t^2},
\end{equation}

\begin{equation}
m_s/m_b = \frac{d (1-d) t^2}{1 + [(1-d)^2 + d^2] t^2}.
\end{equation}

\noindent
Eq. (6) can be inverted to give 
$d = \frac{v}{t} \left( \frac{1 + t^2}{1 + 2 v t}\right)$. 
Substituting this into Eq. (7), one finds that the expressions simplify
to yield a quadratic equation for the
parameter $t = \tan \theta$ in terms of the experimentally known
$v$ and $\ell$:

\begin{equation}
0 = t^2 [5 v (1 - \frac{34}{5} \ell)] + t [1 - 10 \ell] - [\ell/v
+18 \ell v - 3v].
\end{equation}

\noindent
Using $v = 0.035$ and $\ell = 1/50.4$ (these are evaluated at the GUT scale),
one finds that

\begin{equation}
t \equiv  \tan \theta = 0.536, \;\;\; d = 0.081.
\end{equation}

\noindent
Note that the angle $\theta$ is not particularly small. This means
that the Yukawa vectors $a_i$ and $b_i$ are not ``unnaturally" aligned
in family space. This is consistent with the philosophy that there
is no family symmetry. The small parameter in this model is really
$d$. As is evident from Eqs. (6) -- (8), it is the smallness
of $d$ that accounts for the mass hierarchy between the second and 
third families and for the smallness of the mixing between them.
That is, remarkably, it is not a flavor symmetry that produces these 
``flavor" features, but the pattern of $SO(10)$ breaking. Note also
that in the limit $d \rightarrow 0$, the Georgi-Jarlskog relation
$m_s/m_b = \frac{1}{3} m_{\mu}/m_{\tau}$ becomes exact, as can be seen
from Eqs. (7) and (8). In fact, this is how the Georgi-Jarlskog
relation was obtained in the model of \cite{bb}. However, since 
the parameter $d$ is significantly different from zero, there is a
significant deviation from the exact Georgi-Jarlskog prediction
for $m_s$. One finds that

\begin{equation}
m_s/m_b \cong 0.866 (m_s/m_b)_{GJ} \cong 1/58.2.
\end{equation}

\noindent
This is the value at the unification scale. It translates into a 
strange quark mass of
about 137 MeV at 1 GeV, or about 100 MeV at 2 GeV, which is in the 
range given by recent lattice calculations.

\vspace{0.5cm}

\noindent
{\bf Model 2.} The second model is a realization of the model of
\cite{abb} and \cite{ab2} constructed without use of flavor symmetries. 
Again based on $SO(10)$ it
has the following Yukawa superpotential terms:

\begin{equation}
\begin{array}{ccl}
W_{Yukawa} & = & M(\overline{{\bf 16}} {\bf 16}) + 
a_i ( \overline{{\bf 16}} {\bf 16}_i) {\bf 1}_H \\ \\
& + & M' ({\bf 10} {\bf 10}) + b_i ( {\bf 10} {\bf 16}_i ) {\bf 16}_H
\\ \\
& + & c ({\bf 16} {\bf 16}) {\bf 10}_H +
d ( {\bf 16} {\bf 10}) {\bf 16}'_H \\ \\
& + & g_i ( {\bf 16} {\bf 16}_i) {\bf 10}_H {\bf 45}_H {\bf 1}_H/M_G^2.
\end{array}
\end{equation}

\noindent
Here, the spinor Higgs field ${\bf 16}_H$ is supposed to acquire a
VEV in the $SU(5)$-singlet direction, i.e. one that commutes with the
Standard Model group, whereas ${\bf 16}'_H$ is supposed to acquire a VEV
in the Weak-doublet direction \cite{ab1}. 
As in the model of \cite{abb} and \cite{ab2}, the
adjoint Higgs field ${\bf 45}_H$ is supposed to acquire a VEV in
the $B-L$ direction, as needed to solve the doublet-triplet splitting
problem by the Dimopoulos-Wilczek mechanism.
This set of terms can be shown to be the
most general that is consistent with a $Z_3$ symmetry under which
all the quark and lepton multiplets tranform trivially except for the
three ordinary families ${\bf 16}_i$, which all transform non-trivially
and in the same way. 

The form of the mass matrices that result from these Yukawa terms can
be determined by the same kind of reasoning that was used above. 
As before, we can choose our axes to make $a_i = a (0,0,1)$ and
$b_i = b (0, \sin \theta, \cos \theta)$. The coefficient $c_i$ in the
higher-dimension operator will then in general have three non-zero
components.
It is easy to see that integrating out the $\overline{{\bf 16}}$
leads to a mixing of the ${\bf 16}$ and ${\bf 16}_3$, as discussed
before. Similarly, integrating out the superheavy fermions in the 
${\bf 10}$ leads to a mixing of the $SU(5)$ $\overline{{\bf 5}}$'s
in ${\bf 10}$ and in $b_i {\bf 16}_i = 
b(\cos \theta {\bf 16}_3 + \sin \theta {\bf 16}_2)$.
The term with coefficient $c$ then gives 33 entries to all the Dirac
mass matrices $N$, $U$, $D$, and $L$. These contributions are denoted 
``1" in the matrices shown below.
The term with coefficient $d$
gives contributions only to the matrices $D$ and $L$.
This can be seen by looking at the $SU(5)$ decomposition of this
term: $d [{\bf 10}({\bf 16}) \overline{{\bf 5}}( {\bf 10})]
\overline{{\bf 5}}( {\bf 16}'_H)$. This reduces to a term
proportional to ${\bf 10}_3 (\cos \theta \overline{{\bf 5}}_3
+ \sin \theta \overline{{\bf 5}}_2) \overline{{\bf 5}}_H$.
These contributions are denoted by ``$\sigma$" in the matrices below.
Note that these contributions are ``lopsided", giving only
a contribution to $D_{23}$ but not $D_{32}$ and to $L_{32}$ but
not $L_{23}$. Finally, the higher-dimension operator with coefficient
$g_i$ gives terms proportional to ${\bf 16}_3 ( g_3 {\bf 16}_3 +
g_2 {\bf 16}_2 + g_1 {\bf 16}_1)$, which are denoted by  
``$\epsilon_i$" in the matrices below. Altogether, then, the matrices have
the form:

\begin{equation}
\begin{array}{l}
U = \left( \begin{array}{ccc} 
0 & 0 & - \epsilon_1/3 \\ 0 & 0 & - \epsilon_2/3 \\ \epsilon_1/3
&  \epsilon_2/3 & 1 \end{array} \right) M_U, \;\;\; 
D = \left( \begin{array}{ccc} 
0 & 0 & - \epsilon_1/3 \\ 0 & 0 & \sigma s_{\theta} - \epsilon_2/3 
\\ \epsilon_1/3
&  \epsilon_2/3 & 1 + \sigma c_{\theta} \end{array} \right) M_D, \\ \\
L =  \left( \begin{array}{ccc} 
0 & 0 &  \epsilon_1 \\ 0 & 0 & \epsilon_2 \\ - \epsilon_1
& \sigma s_{\theta} - \epsilon_2 & 1 + \sigma c_{\theta} \end{array} 
\right) M_D'
\end{array}
\end{equation}

\noindent
where $s_{\theta} \equiv \sin \theta$, $c_{\theta} \equiv \cos \theta$.
By rotations in the 1-2 planes these can be brought to the forms:

\begin{equation}
\begin{array}{l}
U = \left( \begin{array}{ccc} 
0 & 0 & 0 \\ 0 & 0 & - \epsilon/3 \\ 0
&  \epsilon/3 & 1 \end{array} \right) M_U, \;\;\; 
D \cong \left( \begin{array}{ccc} 
0 & 0 & 0 \\ 0 & 0 & \sigma s_{\theta} - \epsilon_2/3 \\ 0
& \epsilon/3 & 1 + \sigma c_{\theta} \end{array} \right) M_D, \\ \\
L \cong \left( \begin{array}{ccc} 
0 & 0 & 0 \\ 0 & 0 & \epsilon \\ - \epsilon_1
& \sigma s_{\theta} - \epsilon_2 & 1 + \sigma c_{\theta} \end{array} 
\right) M_D'
\end{array}
\end{equation}

\noindent
where $\epsilon \equiv \sqrt{\epsilon_1^2 + \epsilon_2^2}$. 
These matrices go over to those of the model of \cite{abb} in the
limit that $\cos \theta \rightarrow 0$ and $\epsilon_1/\epsilon_2
\rightarrow 0$. As far as the fits to the quark masses and mixings
and the charged lepton masses are concerned, the parameter
$\epsilon_1$ makes very little difference. (It does, however, make
a contribution to the neutrino mixing parameter $U_{e3}$.)
The presence of the $\cos \theta$ and $\sin \theta$ in these matrices
is important, on the other hand, since it introduces an additional
parameter compared to the model of \cite{abb}. There it was
found that an excellent fit was obtained with $\sigma \cong 1.7$
and $\epsilon \cong 0.14$. Here, because of the additional parameter
$\theta$ a slightly better fit is possible. We find the best
fit to be $\sigma \cong 1.6$, $\epsilon \cong 0.15$, and 
$\cos \theta \cong 0.13$. 

Essentially, then, the model is the same as that in \cite{abb}, although
slightly less predictive. It has the same important feature of 
highly ``lopsided" mass matrices $D$ and $L$, i.e. $D_{32} \ll D_{23}
\sim 1$ and $L_{23} \ll L_{32} \sim 1$. This, as emphasized in \cite{abb},
gives a natural explanation of why the atmospheric neutrino mixing
($U_{\mu 3}$) is so large. Other important features are the natural
explanation of the Georgi-Jarlskog factor and of the fact that
$m_c/m_t \ll m_s/m_b$. The reader is referred to \cite{abb} for
further details.

The interesting thing is that we have succeeded in taking a highly
successful model of quark and lepton masses that exists in the
literature and that was constructed by means of flavor symmetries
which distinguish among the three families, and constructing
a model that is virtually the same without making any use
of flavor symmetries of that kind. 

What we have shown is that in the context of grand unification
it is possible to construct interesting and predictive models
which reproduce the important features of the quark and lepton mass
spectrum, while at the same time
treating all three families on exactly the same footing,
that is, giving them the same quantum numbers. In other words, one can
have flavor without flavor symmetry.

\newpage

\begin{picture}(360,216)
\thinlines
\put(108,108){\line(0,-1){36}}
\put(252,108){\line(0,-1){36}}
\thicklines
\put(36,108){\vector(1,0){36}}
\put(72,108){\line(1,0){72}}
\put(180,108){\vector(-1,0){36}}
\put(180,108){\vector(1,0){36}}
\put(216,108){\line(1,0){72}}
\put(324,108){\vector(-1,0){36}}
\put(176,104){$\times$}
\put(172,96){$M$}
\put(60,116){${\bf 16}_i$}
\put(284,116){${\bf 16}_j$}
\put(140,116){$\overline{{\bf 16}}$}
\put(204,116){${\bf 16}$}
\put(104,112){$a_i$}
\put(248,112){$b_j$}
\put(100,60){$\langle {\bf 45}_H \rangle$}
\put(244,60){$\langle {\bf 10}_H \rangle$}
\put(170,10){{\bf Fig. 1}}
\end{picture}

\vspace{5cm}

\noindent
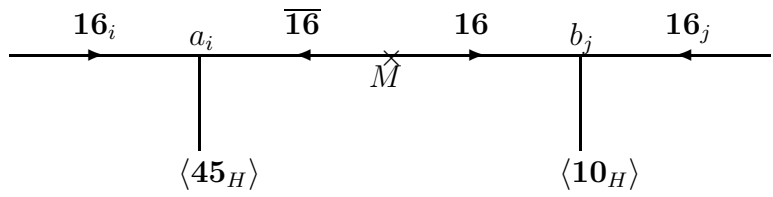
{\bf Fig. 1.} The diagram that leads to the effective operator 
given in Eq. (2).

\end{document}